\documentclass[a4paper,11pt]{article}
\usepackage[utf8x]{inputenc}

\usepackage{a4wide}
\usepackage{amssymb}
\usepackage{amsmath}
\usepackage{amsthm}
\usepackage[mathscr]{euscript}
\usepackage[normalem]{ulem}
\usepackage[colorlinks=true]{hyperref}
\usepackage[usenames,dvipsnames]{xcolor}
\usepackage{todonotes}
\usepackage{appendix}
\usetikzlibrary{arrows,angles}
\usetikzlibrary{matrix}
\usetikzlibrary{snakes}
\usetikzlibrary{arrows,calc,shapes,decorations.pathreplacing}
\usepackage{tikz-cd}

\numberwithin{equation}{section}

\theoremstyle{definition}

\newtheorem{Definition}{Definition}[section]

\newtheorem{Theorem}[Definition]{Theorem}
\newtheorem{Proposition}[Definition]{Proposition}
\newtheorem{Lemma}[Definition]{Lemma}

\newtheorem{Conjecture}[Definition]{Conjecture}

\theoremstyle{remark}

\newtheorem{Remark}[Definition]{Remark}

\newcommand{\R}{\mathbb R}

\usepackage[xcolor]{changebar}
\cbcolor{blue}
\newcommand{\Ric}{\mathrm{Ric}}

\usepackage[inline]{enumitem}
\newcommand{\enumlabelformat}{\roman}
\newcommand{\enumlabelfont}[1]{#1}
\newlength{\thelabelsep}
\setlist{labelsep=\thelabelsep}
\setlist[enumerate,1]{font=\enumlabelfont,label=(\enumlabelformat*),leftmargin=2.5em}
\setlist[itemize]{leftmargin=2.5em,label=$-$}

\newcounter{inlineenum}
\renewcommand{\theinlineenum}{\enumlabelformat{inlineenum}}

\let\epsilon\varepsilon
\let\phi\varphi

\makeatletter
\let\save@mathaccent\mathaccent
\newcommand*\if@single[3]{%
  \setbox0\hbox{${\mathaccent"0362{#1}}^H$}%
  \setbox2\hbox{${\mathaccent"0362{\kern0pt#1}}^H$}%
  \ifdim\ht0=\ht2 #3\else #2\fi
  }
\newcommand*\rel@kern[1]{\kern#1\dimexpr\macc@kerna}
\newcommand*\widebar[1]{\@ifnextchar^{{\wide@bar{#1}{0}}}{\wide@bar{#1}{1}}}
\newcommand*\wide@bar[2]{\if@single{#1}{\wide@bar@{#1}{#2}{1}}{\wide@bar@{#1}{#2}{2}}}
\newcommand*\wide@bar@[3]{%
  \begingroup
  \def\mathaccent##1##2{%
    \let\mathaccent\save@mathaccent
    \if#32 \let\macc@nucleus\first@char \fi
    \setbox\z@\hbox{$\macc@style{\macc@nucleus}_{}$}%
    \setbox\tw@\hbox{$\macc@style{\macc@nucleus}{}_{}$}%
    \dimen@\wd\tw@
    \advance\dimen@-\wd\z@
    \divide\dimen@ 3
    \@tempdima\wd\tw@
    \advance\@tempdima-\scriptspace
    \divide\@tempdima 10
    \advance\dimen@-\@tempdima
    \ifdim\dimen@>\z@ \dimen@0pt\fi
    \rel@kern{0.6}\kern-\dimen@
    \if#31
      \overline{\rel@kern{-0.6}\kern\dimen@\macc@nucleus\rel@kern{0.4}\kern\dimen@}%
      \advance\dimen@0.4\dimexpr\macc@kerna
      \let\final@kern#2%
      \ifdim\dimen@<\z@ \let\final@kern1\fi
      \if\final@kern1 \kern-\dimen@\fi
    \else
      \overline{\rel@kern{-0.6}\kern\dimen@#1}%
    \fi
  }%
  \macc@depth\@ne
  \let\math@bgroup\@empty \let\math@egroup\macc@set@skewchar
  \mathsurround\z@ \frozen@everymath{\mathgroup\macc@group\relax}%
  \macc@set@skewchar\relax
  \let\mathaccentV\macc@nested@a
  \if#31
    \macc@nested@a\relax111{#1}%
  \else
    \def\gobble@till@marker##1\endmarker{}%
    \futurelet\first@char\gobble@till@marker#1\endmarker
    \ifcat\noexpand\first@char A\else
      \def\first@char{}%
    \fi
    \macc@nested@a\relax111{\first@char}%
  \fi
  \endgroup
}
\makeatother

\newcommand{\pd}{\partial}

\title{Vacuum cosmological spacetimes without CMC Cauchy surfaces}

\author{Eric Ling\footnote{Copenhagen Centre for Geometry and Topology (GeoTop), Department of Mathematical Sciences, University of Copenhagen, DK-2100 Copenhagen, Denmark, el@math.ku.dk}\\Argam Ohanyan\footnote{Department of Mathematics, University of Vienna, Oskar-Morgenstern-Platz 1, 1090 Wien, Austria, \newline argam.ohanyan@univie.ac.at}
}
\begin{document}

\date{\today}


\maketitle

\begin{abstract}

In this article, we extend a construction of \cite{chrusciel2005initial} to obtain a large class of vacuum cosmological spacetimes that do not contain any CMC Cauchy surfaces.\ The allowed spatial topologies for these examples are of the form $M \# M$, where $M$ is any closed, connected, oriented, irreducible $3$-manifold which is not spherical.\ This complements the recent results of \cite{LingOhanyan2024}, where, instead of initial data methods, global spacetime gluing arguments were used. The study of such examples is sure to yield insight into Bartnik's cosmological splitting conjecture \cite{bartnik1988remarks}.
\vspace{1em}

\medskip


\end{abstract}

\section{Introduction}

Let us discuss the context in which the need to study cosmological spacetimes without CMC Cauchy surfaces arises.\ We begin by recalling the cosmological version of the Hawking--Penrose singularity theorem \cite{hawking1970singularities} (see also \cite{HawkingEllis1973}), where in this context \emph{cosmological} means globally hyperbolic with compact Cauchy surfaces.

\begin{Theorem}[Cosmological Hawking--Penrose singularity theorem] Let $(\Tilde{M},\Tilde{g})$ be a cosmological spacetime of dimension $\dim \tilde{M} \geq 3$ satisfying the strong energy condition, i.e., $\Ric(v,v) \geq 0$ for all $v \in T\tilde{M}$ timelike. Suppose $(\tilde{M},\tilde{g})$ satisfies the generic condition, i.e., for each inextendible causal geodesic $\gamma$ there exists a parameter $t$ for which $\gamma$ is defined such that the tidal force operator $T_{\gamma(t)}\tilde{M} \to T_{\gamma(t)}\tilde{M}$ defined by $v \mapsto R(v,\gamma'(t))\gamma'(t)$ is not the zero operator. Then there are incomplete causal geodesics in $\tilde{M}$.
\end{Theorem}

Clearly, this result is wrong without the assumption of genericity. Indeed, let $(S,g)$ be any compact Riemannian manifold of dimension $\dim S \geq 2$ with $\Ric_g \geq 0$, then the product spacetime $(\R \times S, -dt^2 + g)$ is cosmological, satisfies the strong energy condition, and is geodesically complete. In this example, genericity is violated along the vertical timelike lines $t \mapsto (t,x_0)$ for any $x_0 \in S$.

As is often the case in global geometric results which involve curvature assumptions, one may now wonder whether the cosmological Hawking--Penrose singularity theorem is rigid in its assumption of genericity, i.e., whether every causally geodesically complete cosmological spacetime satisfying the strong energy condition is isometric to a product spacetime of the form $(\R \times S, -dt^2 + g)$, with $(S,g)$ a compact Riemannian manifold of nonnegative Ricci curvature. A slightly stronger result (assuming only timelike geodesic completeness) was conjectured by Bartnik \cite{bartnik1988remarks} and is one of the most significant open problems in spacetime geometry and mathematical General Relativity to date:

\begin{Conjecture}[Bartnik's cosmological splitting conjecture] Suppose $(\tilde{M},\tilde{g})$ is a cosmological spacetime of dimension $\dim \tilde{M} \geq 2$ satisfying the strong energy condition. If $(\tilde{M},\tilde{g})$ is timelike geodesically complete, then it is isometric to a product spacetime $(\R \times S, -dt^2 + g)$, where $(S,g)$ is a compact Riemannian manifold of nonnegative Ricci curvature.
\end{Conjecture}

Bartnik's conjecture is known to hold under various additional assumptions, we refer to \cite{galloway2019existence} for a detailed discussion of these developments. In particular, let us mention that it holds under the stronger assumption of nonpositive timelike sectional curvature \cite{galloway2018existence}, which allows for powerful timelike geometric comparison results such as triangle or hinge comparison (see \cite{beran2024equivalence} for a proof of equivalence of these conditions).

Bartnik showed in \cite{bartnik1988remarks} that the claimed splitting in his conjecture is equivalent to the existence of CMC Cauchy surfaces, which in fact is what was shown in \cite{galloway2018existence} under the stronger curvature assumption. We refer to \cite{dilts2017spacetimes} for a discussion on the existence of CMC Cauchy surfaces in cosmological spacetimes. As explained in that reference, most existence results rely on barrier methods, see \cite{Galloway_Ling_Remarks_CMC} for an application. Moreover, Bartnik constructed a timelike geodesically incomplete cosmological spacetime which satisfies the strong energy condition and which contains no CMC Cauchy surfaces via the following gluing argument: Take one half of maximally extended Schwarzschild and glue\footnote{This gluing is done in the framework of Tolman--Bondi metrics, reducing the problem of gluing spacetimes to the problem of gluing ODE solutions. See \cite{LingOhanyan2024} for a detailed discussion.} it to an FLRW model which is spatially $T^3$, then extend it to the Schwarzschild event horizon and attach there a time-inverted copy of the same spacetime. One can show that the resulting spacetime has no CMC Cauchy surfaces either via a topological argument, using $3$-manifold topology, or using Hawking's cosmological singularity theorem. This construction was recently extended \cite{LingOhanyan2024} to include more general topologies on the spatial slices of FLRW, and even allowed for different slices on the two sides. If the spatial slices on the two sides agree and are $3$-dimensional, then one can still argue in the same topological way as Bartnik by using the positive resolution of the surface subgroup conjecture \cite{kahn2012immersing}. In all other cases (varying sides, higher dimensions), one argues via the Hawking singularity theorem.

It should be noted that Bartnik's example (as well as the generalizations in \cite{LingOhanyan2024}) are non-vacuum. A vacuum analogue was obtained by Chru{\'s}ciel, Isenberg, and Pollack in \cite{chrusciel2005initial}, where, instead of spacetime gluing, the authors used their results on the gluing of initial data. Essentially, they glued suitable initial data of the form $(T^3,g,K)$ to its ``reverse" $(T^3,g,-K)$ in such a way that in the resulting initial data $(T^3 \# T^3, \bar{g}, \bar{K})$, the metric was symmetric under a reflection across the middle of the connecting neck, while the second fundamental form switches signs. The same topological arguments as the ones used by Bartnik \cite{bartnik1988remarks} apply, showing that the maximal globally hyperbolic development of this data does not contain any CMC Cauchy surfaces.

Our contribution in this paper is to extend the example of \cite{chrusciel2005initial} by generalizing $T^3$ to an arbitrary compact, connected, oriented, irreducible, non-spherical $3$-manifold $M$. Due to the topological nature of the argument, this generalizes the case of $3$ space dimensions and symmetric sides in the gluing, but not the more general situations considered in \cite{LingOhanyan2024}.

\section{The main result}

Let $M$ be a closed, connected, oriented 3-manifold. $M$ is called \emph{prime} if it cannot be written as a nontrivial connected sum of two 3-manifolds. $M$ is called \emph{irreducible} if every embedded 2-sphere bounds a 3-ball. Clearly, $M$ is prime whenever its irreducible. Conversely, if $M$ is prime, then either $M$ is irreducible or it is topologically $\mathbb{S}^1 \times \mathbb{S}^2$ \cite[Lem.\ 3.13]{Hempel3manifolds}. Suppose $M$ is irreducible. Either $M$ has finite fundamental group or not. In the former case, $M$ is topologically a spherical space by the positive resolution of the elliptization conjecture.\footnote{By a \emph{spherical space}, we mean any quotient of the 3-sphere $\mathbb{S}^3/\Gamma$ by a discrete subgroup $\Gamma \subset \text{SO}(4)$ acting freely on $\mathbb{S}^3$.} Our main theorem concerns the latter case.

\begin{Theorem}\label{thm: main}
For any closed, connected, oriented, irreducible $3$-manifold $M$ which is not topologically a spherical space, there is a vacuum initial data set on the connected sum $M \# M$  whose maximal globally hyperbolic development does not contain any CMC Cauchy surfaces.
\end{Theorem}

\begin{Remark}
    More generally, Theorem \ref{thm: main} remains true if $M$ is a closed, connected, oriented $3$-manifold whose prime decomposition $M = M_1 \# \dotsb \# M_n$ contains some  factor $M_k$ that is irreducible and not a spherical space.
\end{Remark}

\begin{Remark}
    Observe that the assumption that $M$ is not spherical in Theorem \ref{thm: main} is compatible with the examples constructed in \cite{LingOhanyan2024}. There, the reason spherical spaces were not allowed had to do with the fact that they lead to finite solution intervals when solving the relevant ODE to obtain Tolman--Bondi type metric tensors, making it impossible to glue FLRW spacetimes with spherical spatial slices to Schwarzschild in a globally hyperbolic way. Flat or hyperbolic manifolds do not exhibit this issue.
\end{Remark}

\section{The proof}

A \emph{vacuum initial data set} is a triple $(S,g,K)$, where $S$ is a smooth manifold of dimension $\dim S \geq 2$, $g$ is a Riemannian metric on $S$, and $K$ is a symmetric $(0,2)$ tensor on $S$ such that $g$ and $K$ satisfy the vacuum Einstein constraint equations,\footnote{We do not consider the cosmological constant in this paper, i.e., $\Lambda = 0$.}
\begin{align}
R + H^2 - K_{ij}K^{ij} \,&=\, 0
\\
\nabla_i(K^{ij} - H g^{ij}) \,&=\, 0,
\end{align}
where $R$ is the scalar curvature of $g$ and $H = \text{tr}_g K$ will be referred to as the mean curvature. If $H$ is constant, then $(S,g,K)$ is called a \emph{constant mean curvature (CMC)} vacuum initial data set.

A pair $(N,Y)$ is called a \emph{Killing initial data (KID)} for an initial data set $(S,g,K)$ if $N$ is a function on $S$ and $Y$ is a vector field on $S$ such that 
\begin{align}
\nabla_{(i}Y_{j)} \,&=\, -NK_{ij}
\\
\nabla_i \nabla_j N \,&=\, N(R_{ij} + H K_{ij} - 2 K_{il}K_{j}^{\:\:l}) - (\mathcal{L}_YK)_{ij},
\end{align} 
where $\mathcal{L}_YK$ is the Lie derivative of $K$ with respect to $Y$ (see \cite[Eqs.\ (5.3) \& (5.4)]{KIDsnongeneric}). If $(\tilde{M}, \tilde{g})$ is the maximal globally hyperbolic vacuum spacetime development of $(S,g,K)$, then KIDs are in one-to-one correspondence with Killing vector fields of $(\tilde{M}, \tilde{g})$ (see \cite{Moncrief1, Moncrief2}).

\medskip

\begin{Lemma}
\label{lemma}
Let $(S,g,K)$ be a CMC vacuum initial data set with $S$ closed, connected and $\dim S \geq 2$. If $(N,Y)$ is a KID for $(S,g,K)$, then $Y$ is a Killing vector field. Moreover, $N$ is constant on $S$, and $N \neq 0$ implies $K=0$.
\end{Lemma}

\proof
It suffices to show the claims about $N$, since either of the options imply that $Y$ is a Killing vector field. We claim that  
\[
\Delta N \,=\, |K|^2 N.
\]
Once this is shown, multiplying this equation by $N$ and integrating by parts shows that $\nabla N = 0$ since $S$ is closed, hence $N$ is constant; the left hand side is manifestly nonpositive and the right hand side is manifestly nonnegative. Since the constancy of $N$ trivially implies $\Delta N = 0$, we either have that $N=0$, or else $K=0$.

Now we prove the claimed equation. Contracting the second KID equation with $g^{ij}$ and using the first constraint equation, we find
\[
\Delta N \,=\, -|K|^2 N - g^{ij}(\mathcal{L}_YK)_{ij}.
\]
We have
\begin{align*}
(\mathcal{L}_YK)_{ij} \,&=\, Y(K_{ij}) - K\big([Y,\pd_i], \pd_j \big) - K\big(\pd_i, [Y,\pd_j]\big).
\end{align*}
CMC initial data implies $Y(H) = 0$ and so $g^{ij}Y(K_{ij}) = -K_{ij}Y(g^{ij}) = K^{ij}Y(g_{ij})$. But
\[
K^{ij}Y(g_{ij}) \,=\, g^{ij}\big(K(\nabla_Y \pd_i, \pd_j) + K(\pd_i, \nabla_Y \pd_j) \big).
\]
Therefore, using the fact that the Levi--Civita connection is torsion-free, we obtain
\[
g^{ij}(\mathcal{L}_YK)_{ij} \,=\, g^{ij}\big(K(\nabla_{\pd_i}Y, \pd_j) + K(\pd_i, \nabla_{\pd_j}Y)\big) \,=\, -2|K|^2N,
\]
where the last equality follows from the first KID equation.
\qed

\medskip

\begin{Proposition}
\label{proposition}
For any closed, connected smooth manifold $S$ of dimension $\dim S \geq 3$, there is a vacuum initial data set $(S,g,K)$ with no nontrivial KIDs.
\end{Proposition}

\proof
By \cite[Thm.\ 2.1, Thm.\ 7.4]{KIDsnongeneric}, there is a Riemannian metric $\hat{g}$ on $S$ such that $\hat{g}$ has no nontrivial local conformal Killing vector fields. Let $\hat{\sigma}$ be a nontrivial traceless and divergence-free tensor with respect to $\hat{g}$. Such tensors exist by general results in \cite{DelayPDE} (in fact, the space of TT tensors is infinite-dimensional; see Remark 9.9 in \cite{DelayPDE}). Hence, the conformal method applies \cite[Sect.\ 4]{Isenbergconformalmethod}, and so there is a vacuum initial data set $(S, g, K)$ such that $g$ is a conformal deformation of $\hat{g}$ and $H = \text{tr}_g K$ is constant, which we may specify to be nonzero. If $(N,Y)$ is a KID for $(S,g,K)$, then Lemma \ref{lemma} implies that $Y$ is a Killing vector field for $(S,g)$, as well as either $N=0$ or else $N$ is a nonzero constant and $K=0$. Since $\hat{g}$ has no nontrivial local conformal Killing vector fields, it follows that $Y=0$. Moreover, $N \neq 0$ and thus $K=0$ is ruled out since we specified $H \neq 0$ in the application of the conformal method. Thus, $N = 0$ and $Y=0$, as claimed. 
\qed

\medskip
\medskip

Let us now come to the proof of Theorem \ref{thm: main} and recall that, in the following, $M$ denotes a closed, connected, $3$-dimensional, oriented, irreducible smooth manifold which is topologically not a spherical space, i.e.,\ not diffeomorphic to a homogeneous space of the form $\mathbb{S}^3/\Gamma$, where $\Gamma \subset SO(4)$ is a discrete subgroup acting freely on $\mathbb{S}^3$.

\medskip
\medskip

\noindent\emph{Proof of Theorem \ref{thm: main}}. 
By Proposition \ref{proposition}, there is a vacuum CMC initial data set $(M,g,K)$ with no nontrivial KIDs.\ As shown in \cite{chrusciel2005initial}, the lack of nontrivial KIDs allows one to perform a gluing procedure on $(M_1,g,K)$ and $(M_2,g,-K)$ with $M_i \approx M$ producing a vacuum initial data set on $M \# M$ such that the metric is symmetric under a reflection across the middle of the connecting neck, while the second fundamental form changes sign under this reflection. Due to this symmetry at the level of the initial data, the resulting maximal globally hyperbolic vacuum spacetime development $(\tilde{M}, \tilde{g})$ admits a time-inverting isometry.

Seeking a contradiction, suppose $(\tilde{M}, \tilde{g})$ contains a CMC Cauchy surface. Then, as in \cite{bartnik1988remarks} (see also \cite{LingOhanyan2024}), using this and the time-inverting isometry produces barriers for the existence of a maximal Cauchy surface $C$ \cite{bartnik1988remarks, Claus_Gerhardt}. The constraint equations imply that $C$ has nonnegative scalar curvature. However, its fundamental group is $\pi_1(C) = \pi_1(M \# M) = \pi_1(M) \ast \pi_1(M)$. Since $M$ is irreducible and not a spherical space, it has infinite fundamental group by the positive resolution of the elliptization conjecture \cite{perelman2002entropy, perelman2003finite, perelman2003ricci}. It follows by the positive resolution of the surface subgroup conjecture \cite{kahn2012immersing} that there is a subgroup of $\pi_1(C)$ which is isomorphic to the fundamental group of a surface of genus $\geq 1$. By \cite[Thm.\ 5.2]{schoen1979existence}, it follows that $C$ is a flat, closed, connected, oriented 3-manifold, but these have been classified and all of them are prime. To see this, recognize that a flat, closed, connected, oriented 3-manifold is a Seifert fibre space (see \cite[Thm.\ 4.3]{Scott3manifolds}). But the only Seifert fibre space which is a nontrivial connected sum is  $\mathbb{R}\mathbb{P}^3 \# \mathbb{R}\mathbb{P}^3$ (see \cite[Prop.\ 1.12]{hatcher2007notes}). But $\mathbb{R}\mathbb{P}^3 \# \mathbb{R}\mathbb{P}^3$ is double covered by $\mathbb{S}^1 \times \mathbb{S}^2$ which does not carry a flat metric \cite[Thm. 3.3.1]{Wolf}. Alternatively, $C$ being $\mathbb{R}\mathbb{P}^3 \# \mathbb{R}\mathbb{P}^3$ topologically would contradict the fact that $M$ is not a spherical space.
\qed

\section{Outlook}

In this article, we generalized the example of a vacuum cosmological spacetime without CMC Cauchy surfaces obtained in \cite{chrusciel2005initial} to initial data on the connected sum $M \# M$ for any closed, connected, oriented, irreducible, non-spherical $3$-manifold $M$. Our topological argument makes use of the positive resolutions of the elliptization conjecture \cite{perelman2002entropy, perelman2003finite, perelman2003ricci} and the surface subgroup conjecture \cite{kahn2012immersing}. Along with examples obtained in \cite{LingOhanyan2024}, we have a plethora of cosmological spacetimes without CMC Cauchy surfaces with varying topologies. But all examples constructed so far have Cauchy surfaces on nonprime 3-manifolds, so it would be interesting to find examples of cosmological spacetimes without CMC Cauchy surfaces whose Cauchy surfaces are prime 3-manifolds.

It should be noted that while the examples obtained by spacetime gluing (\cite{bartnik1988remarks, LingOhanyan2024}) are manifestly timelike and null geodesically incomplete, this is not at all obvious for the ones obtained by initial data gluing (\cite{chrusciel2005initial} and our generalizations in this article). For the one in \cite{chrusciel2005initial}, null geodesic incompleteness has been established in \cite{burkhart2019null} (see also \cite{burkhart2019causal}), while timelike geodesic (in)completeness is entirely open. While the methods employed in \cite{burkhart2019null} do not directly translate to the case of general $3$-manifolds $M$ (as they use the specific form of the $T^3$-metric in coordinates), the methods employed in \cite{burkhart2019causal} show null geodesic incompleteness in any vacuum development of the examples constructed in our main theorem. 

Note that, if Bartnik's splitting conjecture is true, then all of these examples have to be timelike geodesically incomplete.

In general, a lot of work remains to be done to understand the connection between the nonexistence of CMC Cauchy surfaces and timelike resp.\ null geodesic incompleteness so as to obtain greater insight into Bartnik's splitting conjecture. We hope to explore these areas of research further in future works.

\section*{Acknowledgments}

Eric Ling is supported by Carlsberg Foundation CF21-0680 and Danmarks Grundforskningsfond CPH-GEOTOP-DNRF151. Argam Ohanyan is supported in part by the ÖAW-DOC scholarship of the Austrian Academy of Sciences and in part by the project P33594 of the Austrian Science Fund FWF. The authors would like to thank Piotr Chru{\'s}ciel and Greg Galloway for helpful discussions and valuable feedback.

This research was funded in part by the Austrian Science Fund (FWF) [Grant DOI 10.55776/P-33594]. For open access purposes, the authors have applied a CC BY public copyright license to any author accepted manuscript version arising from this submission. 

\addcontentsline{toc}{section}{References}
\bibliographystyle{abbrv}

\end{document}